\documentstyle[seceq,epsf]{ptptex}

\title{The characteristic treatment of black holes}

\author{
Jeffrey {\sc Winicour}
}

\inst{Department of Physics and Astronomy,\\
         University of Pittsburgh, Pittsburgh, PA 15260
}

\recdate{
}

\abst{
The characteristic initial value problem has been successfully implemented as
a robust computational algorithm (the PITT NULL CODE) to evolve
4-dimensional vacuum space-times. It has been applied to the calculation of
gravitational waveforms emitted by black holes and to the event horizon
structure in the merger of black holes. The characteristic code also has
potential application to the binary black hole problem via
Cauchy-characteristic matching.

Because the event horizon is itself a characteristic hypersurface, it can be
analyzed  by characteristic techniques as a stand-alone object. We have
developed an analytic conformal model of null hypersurfaces which gives new
insight into the intrinsic geometry of the pair-of-pants horizon found in the
numerical simulation of the head-on collision of black holes and
into the initially toroidal horizon found in the simulation of a collapsing,
rotating cluster. 

Most studies of black hole formation and merger have been restricted to
axisymmetry. However, axisymmetric horizons, like the Schwarzschild
horizon, are non-generic. When applied to a {\em non}-axisymmetric horizon, the
characteristic approach reveals substantially new features. In particular,
coalescing black holes generically go through a toroidal phase before they
become spherical.  The conformal structure of the event horizon supplies part
of the data for a simulation of the  exterior space-time. This provides a new
way to calculate the post-merger waveforms from a binary black hole inspiral. 
}

\begin{document}

\maketitle

\section{Null Cone Evolution}

A longterm project~\cite{IWW} to develop the pioneering work of
Bondi~\cite{bondi} and Penrose~\cite{null-infinity} into a computational
algorithm for the characteristic initial value problem has recently culminated
in a highly accurate, efficient and robust code - the PITT NULL
CODE~\cite{highp}.   Because the evolution proceeds on a space-time foliation
by null cones which are  are generated by the characteristic rays of the
theory, this approach offers several  advantages for numerical work. I will
describe here the fruits of these investigations relevant to black hole
physics. 

In null cone coordinates, Einstein's equations reduce to propagation
equations along the radial light
rays, which can be integrated in hierarchical order for one variable at a
time. This leads to a highly efficient characteristic marching
algorithm. There is one complex evolution variable which describes the
free degrees of freedom of the gravitational field and four auxiliary
variables. A compactified grid, based upon Penrose's conformal
description of null infinity, removes the necessity of an artificial outgoing
radiation condition and makes possible a rigorous description of geometrical
quantities such as the Bondi mass and news function. The news function supplies
both the true waveform and polarization of the gravitational radiation incident
on a distant antenna.  Furthermore the grid domain is exactly the region in
which waves propagate, which is ideally efficient for the purpose of radiation
studies. Since each null cone extends from the source to null infinity, the
radiation appears immediately with no need for numerical evolution to propagate
it across the grid.  In addition, the growth of a
large redshift offers the bonus of forecasting event horizon formation.

The computational technique of shooting along characteristics is standard
in one spatial dimension but the use of characteristic
hypersurfaces as the underlying foliation for numerical evolution in higher
dimensions is exclusive to relativity~\cite{livr}. The basic approach is
applicable to any of the hyperbolic systems occurring in physics, e.g.
the wave equation, electromagnetic theory and hydrodynamics.

For general relativity, the computational grid is based on coordinates
constructed from a family of outgoing null hypersurfaces emanating from a
worldtube of topology $S^2\times R$. Let $u$ label these hypersurfaces,
$x^A$ ($A=2,3$) be angular coordinates for the null rays and $r$ be a
surface area distance. Then, in the resulting $x^\alpha=(u,r,x^A)$ coordinates,
the metric takes the Bondi-Sachs form~\cite{bondi,sachs}
\begin{eqnarray}
   ds^2= &-& \left(e^{2\beta}{V \over r} -r^2h_{AB}U^AU^B\right)du^2
        -2e^{2\beta}dudr -2r^2 h_{AB}U^Bdudx^A \nonumber \\
         &+& r^2h_{AB}dx^Adx^B,
   \label{eq:bmet}
\end{eqnarray}
where $det(h_{AB})=det(q_{AB})=q(x^A)$, with
$q_{AB}$ a unit sphere metric.	For purposes of including null
infinity as finite grid points, the code uses a compactified
radial coordinate.

The traditional $3+1$ decomposition of space-time used in the
Cauchy formalism is not applicable here because the foliation by null
hypersurfaces of constant $u$ has a degenerate $3$-metric and a null normal.
However, an analogous $2+1$ decomposition can be made on the timelike
worldtube of constant $r$, which has intrinsic metric
\begin{equation}
   {}^{(3)}ds^2=-e^{2\beta}{V \over r}du^2
        +r^2h_{AB}(dx^A-U^Adu)(dx^B-U^Bdu).
\end{equation}
In this form, we can identify $r^2h_{AB}$ as the metric of the surfaces
of constant $u$ which foliate the worldtube, $e^{2\beta}V/r$
as the square of the lapse function and $(-U^A)$ as the shift vector.

A Schwarzschild geometry in outgoing Eddington-Finklestein coordinates is
given by the choice $\beta=0$, $V=r-2m$, $U^A=0$ and $h_{AB}=q_{AB}$. In the
nonspherically symmetric case, in order to computationally treat derivatives
of tensor fields on the sphere, we introduce two stereographic coordinate
patches with a complex unit sphere dyad vector satisfying
$q_{AB} =q_{(A} \bar q_{B)}$.
This allows use of the spin weight {\it eth} operator ~\cite{eth} to express
the covariant derivatives of tensor
fields on the sphere as spin-weighted fields. Our computational {\it eth}
formalism~\cite{competh} accurately calculates covariant derivatives
in spherical coordinates.

The conformal 2-metric $h_{AB}$ can be represented by its dyad component
$J=h_{AB}q^Aq^B /2$. The role of $h_{AB}$ as the null hypersurface data
for the characteristic initial value problem can thus be transferred to
$J$. The Einstein equations impose no constraints so that the complex
function $J$ encodes the two degrees of freedom of the gravitational
field of the null hypersurface.

The evolution algorithm is a computational version of the mixed initial value
problem based upon a worldtube and a null hypersurface~\cite{tam}. Consider a
convex worldtube whose interior contains the sources and whose exterior is an
asymptotically flat region of spacetime. If such a worldtube is sufficiently
large it admits a slicing whose outgoing normal null hypersurfaces extend to
infinity without developing caustics.  Given $J$ on the initial null
hypersurface, Einstein's equations propagate data from the world tube outward
along the outgoing characteristics to determine the exterior space-time.  The
required worldtube data  consists of the conformal 2-geometry of a foliation of
the the worldtube and the mass and angular momentum aspects of the initial
slice. There are constraint equations on the worldtube which are hyperbolic
versions of the elliptic constraints of the Cauchy problem. These constraints
are generalized mass and angular momentum conservation laws. In addition, the
lapse and shift associated with the foliation of the world tube represent gauge
freedom which must be specified.

The worldtube can be shrunk to a nonsingular worldline, in which case the
conservation equations reduce to regularity conditions. The code was first
implemented this way in axisymmetric form~\cite{papa}. However, this has high
computational cost because of the Courant-Friedrich-Lewy condition   which
forces a small time step near the vertex of the null cone. In the 3-dimensional
case, this would make evolution computationally unfeasible on a uniform grid. 
This perhaps can be circumvented by an adaptive grid but at present,
3-dimensional null cone evolution is only feasible in th exterior of a
worldtube. The world tube can be null, as well as timelike, which allows
important application to black holes.

Given worldtube data and initial data, the evolution provides the waveform at
future null infinity ${\cal I}^+$ - first in a super-accelerated frame
determined by the gauge conditions adopted on the world tube and then
converted to an asymptotic inertial frame to give the Bondi news function,
whose real and imaginary parts are the standard plus and cross polarization 
modes. This has been tested to be second order accurate in grid size in a wide
number of test beds including linearized waves and  nonlinear  waves
propagating outside a black hole (which are independently constructed by
solving the Robinson-Trautman equation)~\cite{highp}. 

\section{Black holes}

The Pitt Null Code is a new tool for the accurate simulation of highly curved
space-times. As is historically asked with the discovery of new solutions to
Einstein's equations: ``What can you learn from them?'' The crucial issue in
this question is the physical relevance of the boundary data determining the
space-time, in particular the worldtube data. We would be in great shape
if we knew the right data on a worldtube surrounding the inspiral of a binary
black hole. But the determination of such data is essentially the guts of the
binary black hole problem. 

\subsection{Nonlinear scattering off a black hole}

One simple choice of worldtube is supplied by the ingoing $r=2m$ hypersurface
in a Schwarzschild spacetime (the white hole horizon). The induced worldtube
data automatically satisfies the conservation conditions. With this worldtube
data, we pose the nonlinear version of the classic perturbative
problem~\cite{price} of gravitational wave scattering off a Schwarzschild black
hole by setting initial data on an outgoing null hypersurface extending to
${\cal I}^+$ consisting of an ingoing pulse of compact radial support with
various angular modes.

We have calculated the news function radiated from this system~\cite{highp}. In
the perturbative regime, the news results from the backscattering of the
incoming pulse off the effective potential of the interior Schwarzschild black
hole~\cite{price} and, as expected, scales linearly with the amplitude of the
incoming pulse. However, in the nonlinear regime, the news scales stronger than
linearly with amplitude and the waveform reveals nonlinear generation of
additional modes. In this regime, the mass of the system is dominated by the
incoming pulse, which essentially backscatters off itself in a nonlinear way.
In the extreme nonlinear regime  the total back scattered energy radiated to
${\cal I}^+$ is itself much larger than the mass of the interior black hole. In
dimensionless units the Bondi news $=400$, corresponding to a radiation power
of $10^{13}$ solar masses/sec.

\subsection{Radiation from the capture of matter by a black hole}

We have incorporated a crude hydrodynamic code for a perfect fluid into the
null code. The combined 3-dimensional null code has been tested for stability
and accuracy to verify that nothing breaks, at least in the regime that
hydrodynamical shocks do not form. The results establish the feasibility of a
characteristic {\it matter plus gravity} evolution~\cite{matter}. We used the
code to simulate a localized blob of matter falling into a black hole,
verifying that the motion of the peak of the blob approximated a geodesic and
monitoring the waveform of the emitted gravitational radiation at ${\cal
I}^+$. 

This simulation is a prototype of a neutron star orbiting a black hole. Thus a
refined characteristic hydrodynamic code would open the way to explore an
important astrophysical problem. Recently, such a high resolution
shock-capturing code has been successfully implemented in the null formulation
in a background geometry in the case of spherical symmetry~\cite{toniphil99}
and this code is being generalized to 3-dimensions. We are planning a
collaborative project to combine our characteristic gravitational code with
this characteristic hydro code. 

\subsection{Black hole in a box}

Characteristic evolution can also be based upon a family of {\em ingoing}
null cones with data given on a worldtube at their {\em outer} boundary
and on an initial {\em ingoing} null cone. We implemented such a code to evolve
black holes in the region interior to the worldtube by locating a
marginally trapped surface (MTS) on the ingoing cones and excising the
singular region inside it~\cite{gom97}. This is a characteristic version
of the conventional strategy for evolving black holes by excising the
interior of an apparent horizon, as initially suggested by W.
Unruh\cite{thorn}. The ingoing null code locates the MTS, tracks
it during the evolution and stably excises its interior from the numerical
grid. 

We used this code to simulate a distorted ``black hole in a box''~\cite{wobb}.
Data at the outer worldtube was induced from a Schwarzschild or Kerr spacetime
but the worldtube was allowed to move relative to the stationary trajectories;
i.e. with respect to the grid the worldtube is fixed but the black hole moves
inside it. The initial null data consisted of a pulse of radiation which
subsequently travels outward to the worldtube where it is reflected back toward
the black hole. The approach of the system to equilibrium was monitored by the
area of the MTS, which is equivalent to its Hawking mass. When the worldtube is
stationary (static or rotating in place), the distorted black hole inside
evolves to equilibrium with the boundary. A boost or other motion of the
worldtube with respect to the black hole does not affect this result.  The
marginally trapped surface always reaches equilibrium with the outer boundary,
confirming that the motion of the boundary is ``pure gauge''.

The code essentially runs ``forever'' even when the worldtube wobbles with
respect to the black hole to produce artificial periodic time dependence. 
``Forever'' cannot be rigorously attained in any finite simulation but
we appeal to the characteristic time necessary to obtain accurate
waveforms for the inspiral and merger of two black holes. The inspiral
from a a post-Newtonian orbit at $r=20M$ to the innermost stable orbit at
$6M$ lasts $\approx 10,000M$.  We have successfully evolved an initially
distorted, wobbling black hole for a time of $60,000M$, longer than
needed for a smooth transition from the post-Newtonian regime to merger,
if this success could be duplicated in the ultimate binary black hole
code~\cite{grail}. This capability is important because the coordinates used to
simulate a binary may not become exactly stationary after merger and
ring-down to final equilibrium.

This is the most demanding black hole simulation achieved by any code to date.
Results can be viewed at http://artemis.phyast.pitt.edu/animations. 

\subsection{Cauchy-Characteristic-Matching (CCM)}

The sole weakness of characteristic evolution is its limitation to regions
admitting a nonsingular foliation by global null cones. This problem arises
when focusing by gravity reverses the expansion of the null rays and produces
caustics. Focusing can be magnified by locating the lens far from the
source. In a spacetime with miniscule curvature containing, say, two peanuts,
no global null cones exist if the peanuts are sufficiently far apart
(approximately $10^{10}$ light years). In that case, no matter where the vertex
of a cone is placed, the rays heading toward one of the peanuts would be
refocused. It is not strong curvature by itself but the combination of
curvature and large scale inhomogeneities that makes caustics unavoidable.
In the spherically symmetric gravitational collapse of matter, the light cones
from the center of symmetry extend smoothly to null infinity until the center
enters the event horizon. Global null cones with point vertices exist even for
a binary neutron star with orbital separation less than 5 neutron star radii. 

Given the  appropriate worldtube data for a binary system in its interior,
characteristic  evolution can supply the exterior spacetime and radiated
waveform. But determination of the worldtube data for the complete evolution of
a binary system requires the Cauchy evolution of the interior. CCM is a matched
Cauchy-characteristic evolution designed to tackle such radiation
problems~\cite{Bis90}. The two evolutions are matched across a worldtube, with
the Cauchy domain supplying the boundary values for characteristic evolution
and vice versa. Just as several coordinate patches are necessary to describe a
spacetime with nontrivial topology, an effective attack on the binary black
hole problem is to use CCM to patch together regions of spacetime handled by
different algorithms. 

Because of the singular time dependence of the compactified version of spatial
infinity, a globally compactified approach to the Cauchy problem is not
feasible numerically.  Instead, the grid is terminated at a finite boundary,
where the necessity of an artificial boundary condition, such as the
Sommerfeld condition, leads to strong back reflection in the case of high
asymmetry.  Here the strengths and weaknesses of the Cauchy and
characteristic approaches complement themselves in a fortuitous way. 

The potential advantages of CCM over traditional artificial boundary conditions
are: (1) Accurate waveform and polarization properties at infinity; (2)
Computational efficiency for radiation problems in terms of both the grid
domain and algorithm; (3) Elimination of an artificial outer boundary condition
on the Cauchy problem, which eliminates contamination from back reflection and
clarifies the global initial value problem; and (4) A global picture of the
spacetime exterior to the horizon. These advantages have been realized in the
following two tests of CCM.

First, CCM has been applied to nonlinear scalar waves propagating in a
3-dimensional Euclidean space, matching a spherical null grid to a Cartesian
Cauchy grid~\cite{ccm}.  Performance of the  matching algorithm was compared
with the prime examples of both local and nonlocal radiation boundary
conditions proposed in the computational physics literature. For linear
problems, CCM outperformed all local boundary conditions and was about as
accurate (for the same grid resolution) as the best nonlocal conditions.
However, the computational cost of conventional nonlocal conditions is many
times that of matching so that matching algorithm can be used with finer grids
to yield higher accuracy. For strongly nonlinear problems, matching was {\it
significantly more accurate} than all other methods tested. This is because all
other currently available outer boundary conditions are based on linearizing
the equations in the far field, while CCM consistently takes nonlinearity into
account in both interior and exterior regions.

Second, a two-fold version of CCM has
been used to evolve globally the spherical collapse of a self-gravitating
scalar field onto a black hole~\cite{gom97}. Null evolution on ingoing light
cones, bounded on the inside by a dynamically tracked MTS, are matched at their
outer boundary to the inner boundary of a Cauchy evolution. In turn, the outer
boundary of the Cauchy evolution is matched to an exterior characteristic
evolution extending to infinity.  This study reveals further advantages of CCM
for dealing with black holes. Because the null evolution is extended to
infinity, the appearance of infinite redshifts acts as an automatic mechanism
for stopping the evolution when the event horizon is reached. In addition, the
null approach provides a mechanism for reducing extraneous incoming
radiation from the initial data, thus sharpening the physical model behind the
emitted waveforms.

\begin{figure}
\epsfxsize=3.5in   \epsfysize=3.0in
\centerline{\epsfbox{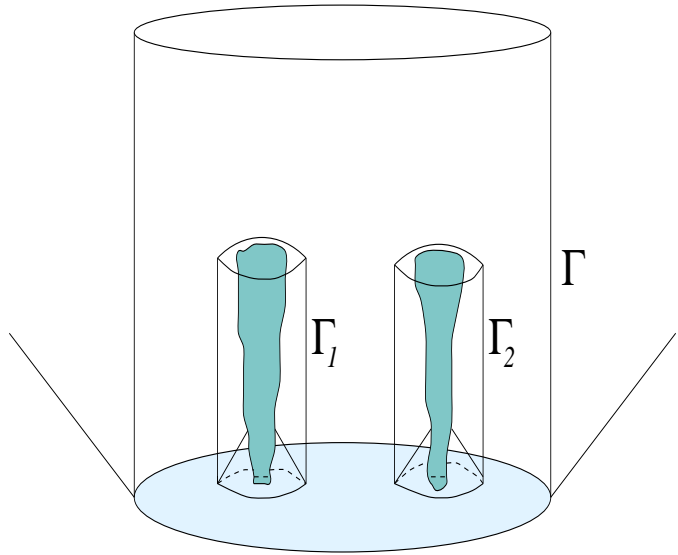}}
\caption{A matching scheme for two orbiting black holes (in a co-rotating
frame which eliminates the major source of time dependence). This
global strategy has been successfully implemented for spherically symmetric
self-gravitating scalar waves evolving in a single black hole spacetime.}
\label{fig1}
\end{figure}

The success of these scalar models suggests an ultimate application of CCM to the
binary black hole problem. Two disjoint characteristic evolutions based upon
ingoing light cones are matched across worldtubes $\Gamma_1$ and $\Gamma_2$ to a
Cauchy evolution of the region between them, as illustrated in Fig. 1. The outer
boundary $\Gamma$ of the Cauchy region is matched to an exterior characteristic
evolution based upon outgoing light cones extending to infinity, where the
waveform is calculated. There are major computational advantages in posing the
Cauchy evolution in a frame co-rotating with the orbiting black holes. Indeed,
such a description may be necessary in order to keep the numerical grid from being
intrinsically twisted. In co-orbiting coordinates the individual holes would be
tracked as they wobble inside the inner matching worldtubes.

All the pieces for such an attack on the binary black hole problem are
presently in place and tested except for the long term stability of
3-dimensional CCM for general relativity. A CCM module incorporating all the
necessary geometric transformations between curved space versions of a
Cartesian Cauchy grid and a spherical null grid has been written and thoroughly
debugged~\cite{vishu}. However, at present there are instabilities of a type
not found in the simpler applications of CCM but similar to those arising from
other choices of Cauchy boundary conditions. 

\section{Characteristic treatment of colliding black holes}

\bigskip

An event horizon is a special type of null hypersurface whose light rays emerge
from an initial caustic-crossover region, where the horizon forms, and then
expand and asymptotically ``hover'' at a finite constant surface area.  In
geometric optics, the chief consideration in the study of wavefronts is the
2-dimensional caustic surface where neighboring rays meet and classically the
intensity would be infinite. The elementary caustics have been classified in
terms of catastrophe optics and, equivalently, in terms of singular maps of
dynamical systems. Because of the inherent structural stability of this
classification, it is unchanged by the tidal distortions produced by spacetime
curvature.  A caustic set consisting of a single point is structurally unstable,
i.e. a small perturbation can produce qualitative changes in the features. This
applies to the point caustic associated with black hole formation in the
Oppenheimer-Snyder model of spherically symmetric collapse. Only the elementary
caustics can play a role in generic horizon formation. 

In contrast to caustics, the classical intensity is finite on the crossover
set, where distinct light rays traced back on the horizon collide. The
crossover set is the dominant structure in horizon formation. Generically, it
is a 2-dimensional surface bounded by a curve of caustics. The generic
properties of crossover surfaces are nonlocal and have not been classified.
After describing the geometry of axisymmetric horizons, I will discuss some
generic properties of the simplest type of crossover set and show how they lead
to topological features in the merger of black holes quite different from the
axisymmetric head-on collision.

\subsection{Axisymmetric Horizons}

We have identified the caustic structure of the horizon found in simulations of
the axisymmetric head-on collision of two black holes~\cite{science}. Cusps and
folds are the only structurally stable caustics in the axisymmetric case.
Although the theory of elementary caustics is local, we have used it to piece
together a global model of the ``trousers'' shaped horizon obtained in the
numerical simulations. Some surprising spacetime features emerge. As
schematically illustrated in Fig. 2, the light rays generating the horizon
originate on a spacelike seam ${\cal X}$ (a set of crossover points) running
along the inside of the trouser legs. The trousers are bowlegged with no
special sharpness at the crotch. For black holes formed at a finite time by the
collapse of matter, the crossover seam extends around the bottom of each
trouser and slightly up the outer side where it becomes asymptotically null and
terminates at a cusp, where neighboring rays focus. 

\begin{figure}
\epsfysize = 12.0cm
\centerline{\epsfbox{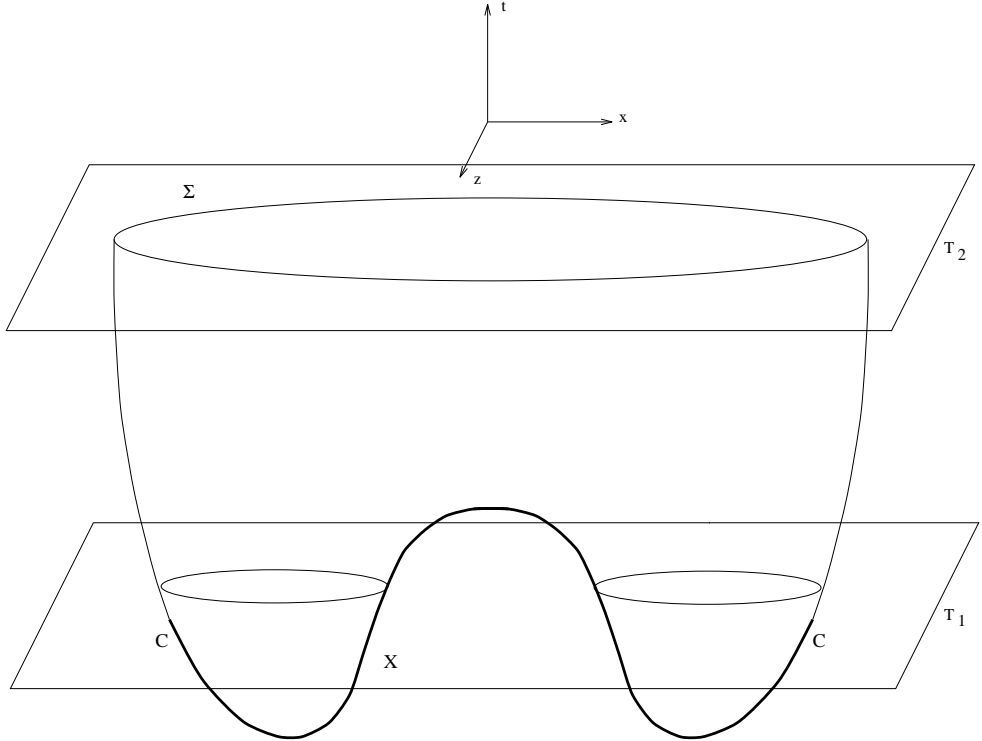}}
\caption{Spacetime picture of a pair-of-pants horizon emerging from
the (thicker shaded) crossover points ${\cal X}$ bounded by cusps $C$.}
\label{fig2}
\end{figure}

For vacuum black holes, the axisymmetry implies the vanishing of focusing along
the umbilical (shear-free) rays on the axis of symmetry. As a result, the
individual black holes in the head-on collision are eternal. The trouser legs
extend forever into the past, although their cross-sectional area becomes
vanishingly small (in the past) as all rays except those on the axis leave the
horizon. Asymptotically, the initial black holes holes are spherical due to the
umbilical nature of the axis.

Surprisingly, a trousers shaped horizon also appears in the axisymmetric
gravitational collapse of a rotating cluster~\cite{torus,toroid}, except now
the rotational symmetry axis in Fig. 2 must be identified differently. In the
head-on collision, the symmetry axis lies in the $x$-direction of Fig. 2. After
replacing the $t$-direction in the figure by the suppressed spatial dimension
(the $y$-direction), a rotation about the $x$-axis produces two spherical black
holes at time $T_1$. Also, in the head-on collision, the crossover points
${\cal X}$ in Fig. 2 lie on the rotation axis so that they generate a crossover
line under rotation. 

For the horizon formed by the rotating cluster, the symmetry axis in Fig. 2
lies the $z$-direction. Now replacing the $t$-direction in the figure by the
suppressed $y$-direction, a rotation about the  symmetry axis produces a
single toroidal black hole at time $T_1$. Furthermore, the crossover points
generate a disc with a circular boundary of cusps. In both models, at the
later time $T_2$ the horizon consists of a single spherical black hole $\Sigma$,
which in one case results from the collision of two black holes and in the
other from the hole in the torus closing up.

The discovery of temporarily toroidal black holes raised the concern~\cite{jac}
of a potential mechanism for violating the topological censorship theorem,
which requires that any two causal curves extending from past to future null
infinity be deformable into each other~\cite{fsw}. The key issue is whether two
twins starting at the same spacetime point outside the horizon could travel to
another spacetime point outside the horizon by homotopically inequivalent
causal paths. A light ray traveling from the infinite past through the hole in
the torus and back out to future null infinity would not be causally deformable
to a light ray that altogether skirts the horizon if the hole in the torus were
{\it too long lived}. However, because the intersection of two null
hypersurfaces is spacelike, the crossover set ${\cal X}$ in Fig. 2 must also be
spacelike so that the the hole closes up superluminally. Consequently, a causal
curve passing inside the hole at a given time can be slipped below the bottom
of a trouser leg to yield a causal curve lying entirely outside the hole in the
torus.

\subsection{The conformal model of black hole coalescence}

Remarkably, the horizon geometries found in the axisymmetric numerical
simulations can be independently obtained from an analytic model based upon the
conformal rescaling of a flat space null hypersurface~\cite{ndata}. For a null
hypersurface emanating from a prolate spheroid (cigar shaped ellipsoid of
revolution), the conformal model reproduces the pair-of-pants found in the
head-on collision of black holes. For an oblate spheroid (pancake shaped
ellipsoid of revolution), it yields the temporarily toroidal horizon found in
the collapse of a rotating cluster. But the model is not confined to
axisymmetry and reveals features of the axisymmetric head-on collision which
are not generic. In a generic black hole merger there is always a toroidal
phase~\cite{asym,siino2}. The analytic nature of the model also provides insight
into the saddle shape geometry at the crotch of the trousers and implications
for the hoop conjecture.

The conformal model treats the horizon in stand-alone fashion as a
3-dimensional manifold endowed with a degenerate metric $\gamma_{ab}$ and
affine parameter $t$ along its null rays. The metric is obtained from the
conformal mapping $\gamma_{ab}=\Omega^2 \hat \gamma_{ab}$ of the intrinsic
metric $\hat \gamma_{ab}$ of a flat space null hypersurface emanating from a
convex surface ${\cal S}_0$ embedded at constant time in Minkowski space.  The
horizon is identified with the null hypersurface formed by the inner branch of
the boundary of the past of ${\cal S}_0$, and its extension into to the future
to ${\cal I}^+$. The flat space null hypersurface expands forever as its affine
parameter $\hat t$ (given by Minkowski time) increases but the conformal factor
is chosen to stop the expansion so that the cross-sectional area of the black
hole approaches a finite limit in the future. At the same time, the Raychaudhuri
equation (which governs the growth of surface area) forces a nonlinear relation
between the affine parameters $t$ and $\hat t$ which introduces the nontrivial
topology of the affine slices of the black hole horizon.

The {\em number} of black holes or the topology of a black hole at a given time
is not conventionally defined in terms of such a stand-alone model but in terms
of the number and type of disjoint intersections with a Cauchy hypersurface. In
the conformal model of a black hole collision the notion of ``two holes'' and
their merger arises intrinsically from the affine foliation of the horizon. It
is the relative distortion between the affine parameters $t$ and $\hat t$
brought about by curved space focusing which gives rise to the trousers shape.

In the conformal model, the caustic-crossover set for the black hole horizon
inherits properties from its flat space counterpart. The generic features of
the model stem from the structurally stable properties of this set which are
preserved under arbitrary smooth perturbations of ${\cal S}_0$. Classification
of the generic properties of the crossover set ${\cal X}$ is a global problem. 
The simplest case to consider is when ${\cal X}$ is a double-crossover set
consisting purely of points at which precisely two rays intersect. This
includes the horizon formed when ${\cal S}_0$ is ellipsoidal, which we have
used to study black hole coalescence analytically.

Consider then the caustic-crossover structure of the wavefront emanating
backward in time from a smooth convex surface ${\cal S}_0$ embedded at constant
time ${\hat t}=0$ in Minkowski space. The rays tracing out the wavefront
generate a smooth null hypersurface until they reach past endpoints on the
boundary of the past of ${\cal S}_0$. The endpoints consist of a set of caustic
points ${\cal C}$, where neighboring rays focus, and a set of crossover points
${\cal X}$, where distinct null rays collide. We have established the following
generic caustic-crossover properties of such flat space null
hypersurfaces.~\cite{asym}

{\bf  Generic Property 1}: {\it A caustic point is not also a crossover point.}

{\bf  Generic Property 2}: {\it Considered as a subset of Minkowski space, the
double-crossover set is a smooth, open, spacelike 2-surface.}

{\bf Generic Property 3}: {\it In the absence of triple or higher order
crossovers,  the caustic set forms a compact boundary to the crossover set
(considered as a subset of the horizon). The tangent space of the crossover set
joins continuously to the tangent space of the null portion of the horizon at
this caustic boundary.}

{\bf Generic Property 4}: {\it A crossover point lies at the intersection of at
most four rays.}

These generic properties are violated in the case of a spherical wavefront,
where the crossover set and caustic set both degenerate to a common point. They
are also violated in the prolate spheroidal case, where the crossover set is a
curve of non-generic caustics, which is bounded at each end by a non-degenerate
cusp type caustic \cite{toroid}. These generic properties are satisfied when
${\cal S}_0$ is a triaxial ellipsoid (no degeneracies in the lengths of the
major axes). For this case, the crossover set is smooth and
connected, consists purely of double crossover points and goes asymptotically
null at an elliptical caustic boundary where it joins smoothly to the null
hypersurface. More complicated examples would allow higher order crossovers.
However, Property 4 limits the complexity of a generic crossover set arising in
Minkowski space from a smooth, convex surface ${\cal S}_0$.

The conformal construction of the curved space event horizon preserves the
structure of the underlying flat space crossover set if appropriate
identifications can be made. This is the case in the triaxial ellipsoid model
because reflection symmetry of the ellipsoid supplies the consistency
conditions for the conformally transformed metric to be single valued at the
crossover points. The generic properties of an arbitrary event horizon are
important features of black hole physics. Certain of the flat space properties
generalize easily to curved space, e.g. the spacelike nature of the crossover
set. However, for those properties established using specifically Euclidean
constructions, e.g. Generic Property 4, the generalization is not obvious. 

The black hole event horizon associated with a triaxial ellipsoid  reveals
new features which are not seen in the axisymmetric head-on collision of two
black holes corresponding to the degenerate case of a prolate spheroid. If the
degeneracy is slightly broken, the physical picture of a black hole collision
remains.  The individual black holes form with spherical topology but, as
illustrated in Fig. 3, as the holes approach, tidal distortion produces sharp
pincers just prior to merger. At merger, the pincers join to form a single
temporarily toroidal black hole, as illustrated in Fig. 4. The inner hole of
the torus subsequently closes up (superluminally) to produce first a peanut
shaped black hole and finally a spherical black hole. In the degenerate
axisymmetric limit, the pincers reduce to a point so that the individual holes
have teardrop shape and they merge without a toroidal phase.  Details of this
merger can be viewed at http://artemis.phyast.pitt.edu/animations. 

\begin{figure}
\epsfxsize = 8.0cm   \epsfysize = 6.0cm
\centerline{\epsfbox{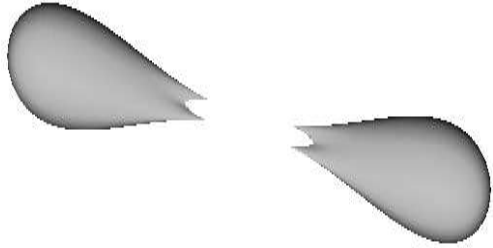}}
\caption{Prelude to a generic collision of black holes.}
\vspace*{10pt}
\label{fig3}
\end{figure}

\begin{figure}
\epsfxsize = 8.0cm   \epsfysize = 6.0cm
\centerline{\epsfbox{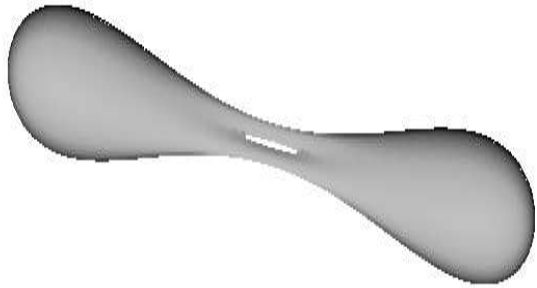}}
\caption{Temporarily toroidal black hole produced by the merger of two black
holes.}
\vspace*{10pt}
\label{fig4}
\end{figure}

\section{Waveforms from coalescing black holes}

A prime application of the conformal horizon model is the calculation of the
waveform emitted by coalescing black holes using the null code. The model
supplies the data on the horizon necessary for an evolution of the exterior
spacetime. The evolution is carried out along a family of ingoing null
hypersurfaces which intersect the horizon in topological spheres. The evolution
is restricted to the period from merger to ringdown, for otherwise the ingoing
null hypersurfaces would intersect the horizon in disjoint pieces. In the
strategy being pursued, the remaining data necessary for the evolution is the
conformal geometry of the {\it final} ingoing null hypersurface in the family,
which is taken to approximate ${\cal I}^+$; i.e. the missing piece of data is
essentially the outgoing waveform! The evolution then proceeds {\it backward}
in time to determine the exterior spacetime in the post-merger era. Since the
outgoing waveform is a necessary piece of data, it might seem that this is a
circular approach. However, as explained below, a variant of this strategy
supplies the physically correct waveform.

The evolution is an implementation of the construction of a vacuum space-time
based upon the characteristic initial value problem posed on an intersecting
pair of null hypersurfaces~\cite{sachsdn,haywsn,haywdn}. Here one of the null
hypersurfaces is the event horizon ${\cal H}$. The other is the ingoing null
hypersurface $J^+$ which intersects ${\cal H}$ in the topologically spherical
surface ${\cal S}_0$. The conformal model provides the geometry of ${\cal S}_0$
and induces the necessary data on the horizon for, say, a black hole collision,
in terms of the geometry of a flat space ellipsoidal null hypersurface. We
locate ${\cal S}_0$ as a cross-section of the event horizon at a late
quasi-stationary time so that the ingoing null hypersurface $J^+$ approximates
${\cal I}^+$, as illustrated in Fig. 5. Part of the horizon data is a quantity,
called the twist \cite{haywsn,haywdn}, which is analogous to the extrinsic
curvature in the Cauchy problem. The twist need only be specified on ${\cal
S}_0$ and then a component of Einstein's equations propagates it along the
generators of the horizon by a simple ordinary differential equation. The
initial value of the twist can be specified in terms of its asymptotic value in
the final Kerr black hole.

The data is completed by specifying the conformal geometry of $J^+$ to the past
of ${\cal S}_0$.  There are existence theorems for solutions to this double
null initial value problem~\cite{hagenseifert77,helmut81a,helmut81b}. Although
global issues remain unresolved, these theorems guarantee existence in some
neighborhood of the initial null hypersurfaces.  Thus, referring to Fig. 5,
the above data determines a solution of the vacuum equations in the
a domain of dependence $D^-$ to the past of $S_0$. (In the figure we assume
the absence of singularities in the past of $S_0$.)

This double null version of the characteristic initial value problem is
equivalent to the world tube - null cone problem~\cite{tam} in the case where
the world tube is null. This is precisely the problem for which the PITT code
provides a stable evolution which is highly accurate even when run on a single
processor.  The code also determines the space-time in a subregion of the
domain of dependence $D^-$, thus providing the local existence of a space-time
satisfying Einstein's equations in the finite difference approximation.
Ideally, the numerical evolution would extend throughout the domain of
dependence $D^-$ but that is more problematical. The evolution algorithm
requires the foliation of $D^-$ by a one parameter family of null hypersurfaces
$J_v$. However, referring to Fig. 5, such a foliation becomes singular at
$J_M$ in the portion of the horizon corresponding to the black hole merger.
Thus only the post-merger space-time is determined by the evolution. This
problem is due to a coordinate singularity, not a physical singularity, and
from a mathematical point-of-view the space-time is extendable to earlier
times; but just how much earlier cannot be answered by the numerical
evolution.  Irregardless, the conformal model of the horizon for a black hole
collision can be used as characteristic initial data to construct a vacuum
space-time  covering a very interesting nonlinear domain from merger to
ringdown.

To the extent that $S_0$ lies in the late quasi-stationary era of the horizon,
the null hypersurface $J^+$ approximates future null infinity. So, for a
horizon describing a binary black hole, the physically appropriate
characteristic data on $J^+$ describes the sought after outgoing radiation
emitted during the black hole merger. However, unlike the Cauchy problem, the
data on a null hypersurface for a solution of Einstein's equations can be posed
in a constraint free manner and any such data leads to a vacuum solution.  This
allows us to compute the merger-ringdown waveform in the following manner.
Since the outgoing waveform is a necessary but arbitrary part of the data, we
first proceed by setting the outgoing waveform to zero. Next we evolve backward
in time to calculate the incoming radiation entering from ${\cal I}^-$. There
is no problem setting up the null data on $J^+$ to be asymptotically flat so
that ${\cal I}^-$ exists, at least for some finite evolution time. This
radiation entering from ${\cal I}^-$ is eventually absorbed by the black hole.
The idea is to cancel it out by adding the right amount of outgoing radiation
on $J^+$. If this calculation were carried out in the perturbative regime of a
a Schwarzschild or Kerr background geometry, as in the close
approximation~\cite{jorge}, this can be accomplished by taking advantage of the
time reflection symmetry.

In order to illustrate the simplicity of this approach, consider the null
evolution of a perturbation of a Schwarzschild background. The numerical
evolution provides the complex metric quantity $J(v,r,\theta,\phi)$ which
determines the conformal null geometry of the ingoing foliation. The incoming
news function $N_{in}(v,\theta,\phi)$ is obtained from the asymptotic
dependence of $J$ near ${\cal I}^+$. By construction, there is no outgoing
radiation in this perturbation. Application of the time reflection symmetry of
the Schwarzschild background provides the alternative perturbation given by
\begin{equation} 
      J_{out}(-t+r^{*},r,\theta,\phi)=J_{in}(t+r^{*},r,\theta,\phi)
\label{eq:subst}
\end{equation} 
where $v=t+r^{*}$ and $u=t-r^{*}$ are advanced and retarded
Schwarzschild times. 

The perturbation $J_{out}$ contains outgoing radiation but no incoming
radiation. Thus the perturbation $J_{out}$ satisfies appropriate boundary
conditions for the collision of two black holes, with no incoming fields at
least in the post-merger period. The corresponding news function $N_{out}$ for
the outgoing waveform is obtained from $N_{in}$ (supplied by the simulation) by
a substitution analogous to Eq. (\ref{eq:subst}).

In the same perturbative regime as the close approximation, this numerical
procedure supplies the  outgoing waveform under the physically appropriate
condition of no ingoing radiation.  More generally, beyond the perturbative
regime, the appropriate outgoing waveform can be obtained by a more complicated
inverse scattering procedure which minimizes the incoming radiation with
respect to variations of the null data on $J^+$. 

\begin{figure}
\epsfxsize=4in
\centerline{\epsfbox{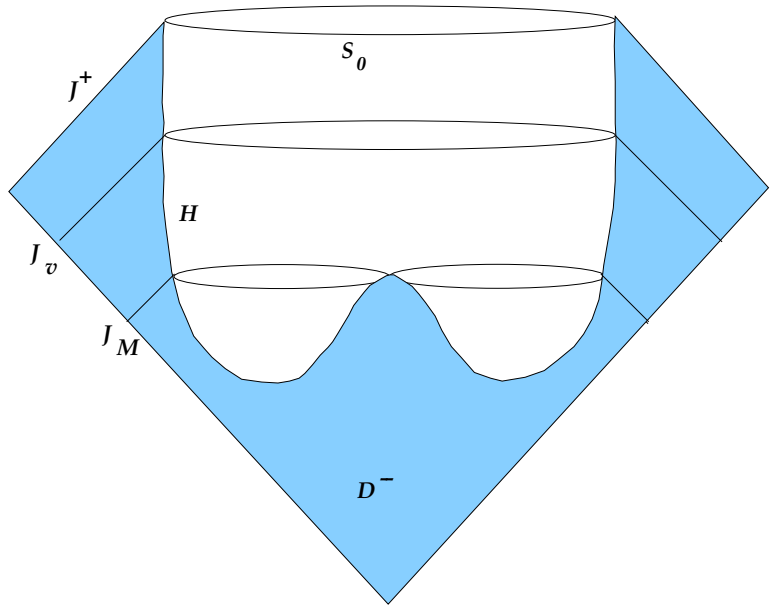}}
\caption{A portion of the space-time prior to the coalescence of two black
holes. The parameter $v$ labels the advanced time on a family of incoming
null hypersurfaces. $D^-$ is the domain  of dependence of characteristic data
given on the event horizon ${\cal H}$ and on $J^+$.}
\label{fig5}
\end{figure}

Thus the conformal horizon model combined with the null evolution code offers
a new way to calculate the merger-ringdown waveform from coalescing black holes.
Because this is an unexplored area of binary black hole physics we are
beginning this study in the simple case of a head-on collision, where the close
approximation waveform has been calculated. This will provide some preliminary
physical checks for extending the work into the nonlinear and nonaxisymmetric
case where inspiraling black holes can be treated.  Preliminary calculations
for the head-on-collision show that at late times the waveform is entirely
quadrupole ($\ell =2$) in agreement with the close approximation, but that a
strong $\ell =4$ mode exists just after merger.

It is interesting to note that turned upside down (time reversed), Fig. 5 
describes a white hole fission. In that case, setting the null data to zero on
$J_v$ corresponds to no ingoing radiation and the numerical calculation
supplies the outgoing waveform from the fission with the physically correct
boundary condition.. It is this white hole fission waveform that is used to
generate the physically correct waveform for a black hole merger, but the
solution to the fission problem is of itself at least of academic interest.

\section*{Acknowledgments}

The results reported here are cumulative of years of productive work by members
of the Pittsburgh numerical relativity group. I am especially grateful to
Roberto G\'omez who supplied the continuity and guidance critical to bringing
the PITT null code to fruition. The graphics were produced by Sascha Husa, Luis
Lehner and Joel Welling. The work was supported by NSF grants PHY 9510895, PHY
9800731 and NSF INT 9515257. Computer time has been provided by the Pittsburgh
Supercomputing Center and the San Diego Supercomputing Center.


\begin{thebibliography}{99}

\bibitem{IWW}
R.~A. Isaacson, J.~S. Welling, and J. Winicour, \JL {J.\ Math.\ Phys.,
24,1983,1824}.

\bibitem{bondi}
M. van~der Burg, H. Bondi, and A. Metzner, \JL {Proc.\ R.\ Soc. \ London,
A269,1962,21}.

\bibitem{null-infinity}
R. Penrose, \PRL {10,1963,66}.

\bibitem{highp} N.~T. Bishop, R.~G\'{o}mez, L. Lehner, M. Maharaj and
J. Winicour, \PR {6,1997,6298}.

\bibitem{livr} J. Winicour, Living Reviews (1998),
http://www.livingreviews.org/

\bibitem{sachs}
R. Sachs, \JL {Proc. \ R. \ Soc.  London ,A270,1962,103}.

\bibitem{eth} E. T. Newman and R. Penrose, \JL { J. \ Math. \ Phys.,7,1966,
863}.

\bibitem{competh} R. G\'{o}mez, L. Lehner, P. Papadopoulos and J. Winicour,
\JL {Class. \ Quantum \ Grav.,14,1997,977}.

\bibitem{tam} L. A. Tamburino and J. Winicour, \PR {150,1966,1039}.

\bibitem{papa}
R. G\'omez, P. Papadopoulos, and J. Winicour, \JL {J.\ Math.\ Phys.,
35,1994,4184}.

\bibitem{price} R.H. Price, \PR {D5,1972,2419}.

\bibitem{matter} N.~T. Bishop, R.~G\'{o}mez, L. Lehner, M. Maharaj and
J. Winicour, \PR {D60,1999,24005}.

\bibitem{toniphil99} 
P. Papadopoulos and J. Font, ``Relativistic hydrodynamics on spacelike
and null surfaces: Formalism and computation of spherically symmetric
spacetimes'' (1999), gr-qc/9902018.

\bibitem{gom97} R. G\'{o}mez, R.  Marsa and J. Winicour,
\PR {D56,1997,6310}.

\bibitem{thorn} See J. Thornburg, \JL {Class. \ Quantum \ Grav.,
4,1987,1119}.

\bibitem{wobb} R.~G\'{o}mez, L. Lehner, R. Marsa, and J. Winicour,
\PR {D57,1997,4778}.

\bibitem{grail} R. G\'{o}mez, L. Lehner, R.  Marsa, J. Winicour and
the BBH Grand Challenge, \PRL {80,1998,3915}.

\bibitem{Bis90} N.~T.Bishop, C. Clarke, and R. d'Inverno, \JL {Class.
\ Quantum. \ Grav.,7,1990,L23}.

\bibitem {ccm}  N.~T. Bishop, R.~G\'omez, P.~R. Holvorcem, R.~A. Matzner,
P. Papadopoulos and J. Winicour,
\JL {J. \ Comp. \ Phys.,136,1997,140}.

\bibitem{vishu} N.~T. Bishop, R. Gomez, L. Lehner, R. Isaacson, B. Szil\'{a}gyi
and J. Winicour, in {\em Black Holes, Gravitational Radiation and the
Universe}, eds. B. R. Iyer and B. Bhawal (Kluwer, Dordrecht, 1998).

\bibitem{science} R.~A. Matzner, H.~E. Seidel, S. L. Shapiro,
L. Smarr, W-M Suen, S. A. Teukolsky, and J. Winicour, \JL {Science,
270,1995,941}.

\bibitem{torus}S.~A. Hughes, C.~R. Keeton, P. Walker, K. Walsh,
S.~L. Shapiro, and S.~A. Teukolsky, \PR {D49,1994,4004}.

\bibitem{toroid} S. Shapiro, S. Teukolsky and J. Winicour, \PR {D52,1995,6982}.

\bibitem{jac} T. Jacobson and S. Venkataramani, \JL {Class. Quantum
Grav.,12,1995,1055}.

\bibitem {fsw} J.~L. Friedman, K. Schleich, and D.~M. Witt,
\PR {71,1993,1486}.

\bibitem{ndata} L. Lehner, N.~T. Bishop, R. G\'{o}mez, B. Szil\'{a}gyi,
and J. Winicour, \PR {D60,1999,44005}.

\bibitem{asym} S. Husa
and J. Winicour, \PR {D60,1999,044005}.

\bibitem{siino2} M. Siino, \PR {D59,1999,064006}.

\bibitem{sachsdn} R.~K. Sachs, \JL {J. \ Math. \ Phys.,3,1962,908}.

\bibitem{haywsn} S.~A. Hayward, \JL {Class. Quantum Grav.,10,1993,773}.

\bibitem{haywdn} S.~A. Hayward, \JL {Class. Quantum Grav.,10,1993,779}.

\bibitem{hagenseifert77}
H. Zum Hagen and H Seifert, \JL {Gen. \ Relat. \ Gravitat.,8,1977,259}.

\bibitem{helmut81a}
H. Friedrich, \JL {Proc. \ Roy. \ Soc. \ London,A375,1981,169}.

\bibitem{helmut81b}
H. Friedrich, \JL {Proc. \ Roy. \ Soc. \ London,A378,1981,401}.

\bibitem{jorge}  R.~H. Price and J. Pullin, \PRL {72,1994,3297}. 

\end{thebibliography}
\end{document}